\documentclass{amsart} 

\usepackage{lscape}
\usepackage{graphicx}
\usepackage[utf8]{inputenc}
\usepackage{amssymb}
\usepackage{amsmath}
\usepackage{siunitx}
\usepackage[caption=false]{subfig}
\usepackage{subfig}
\usepackage{amsaddr}

\def\xR{\mathbb{R}}
\def\x0{\mathbf{0}}
\def\xg{\mathbf{g}}
\def\xi{\mathbf{i}}

\def\xk{\mathbf{k}}
\def\xu{\mathbf{u}}
\def\xe{\mathbf{e}}
\def\xC{\mathbf{C}}
\def\xE{\mathbf{E}}
\def\xI{\mathbf{I}}

\def\xN{\mathbf{N}}

\DeclareMathOperator{\Tr}{Tr}

\begin{document}

\title[Disappearance of stretch-induced wrinkles...]{Disappearance of stretch-induced wrinkles of thin sheets: a study of orthotropic films}


\author{András A. Sipos}
\address{Dept. Mechanics, Materials and Structures \\ Budapest University of Technology and Economics}
\email{siposa@eik.bme.hu}
\author{Eszter Fehér} 
\address{Dept. Mechanics, Materials and Structures \\ Budapest University of Technology and Economics}
\email{fehereszter@gmail.com}
\maketitle

%\address{Department of Mechanics, Materials and Structures, Budapest University of Technology and Economics, Műegyetem rkp. 1-3. K261, Budapest 1111, Hungary}

\begin{abstract}

A recent paper  (Healey \textit{et al., J. Nonlin. Sci.}, 2013, \textbf{23}, 777-805.) predicted the disappearance of the stretch-induced wrinkled pattern of thin, clamped, elastic sheets by numerical simulation of the Föppl-von Kármán equations extended to the finite in-plane strain regime. It has also been revealed that for some aspect ratios of the rectangular domain wrinkles do not occur at all regardless of the applied extension. To verify these predictions we carried out experiments on thin ($20$ \SI{}{\micro\metre} thick adhesive covered), previously prestressed elastomer sheets with different aspect ratios under displacement controlled pull tests. On one hand the the adjustment of the material properties during prestressing is highly advantageous as in targeted strain regime the film becomes substantially linearly elastic (which is far not the case without prestress). On the other hand a significant, non-ignorable orthotropy develops during this first extension. To enable quantitative comparisons we abandoned the assumption about material isotropy inherent in the original model and derived the governing equations for an orthotropic medium. In this way we found good agreement between numerical simulations and experimental data.

Analysis of the negativity of the second Piola-Kirchhoff stress tensor revealed that the critical stretch for a bifurcation point at which the wrinkles disappear must be finite for any aspect ratio. On the contrary there is no such a bound for the aspect ratio as a bifurcation parameter. Physically this manifests as complicated wrinkled patterns with more than one highly wrinkled zones on the surface in case of elongated rectangles. These arrangements have been found both numerically and experimentally. These findings also support the new, finite strain model, since the Föppl-von Kármán equations based on infinitesimal strains do not exhibit such a behavior.

\end{abstract}

\section{Introduction}
\label{Intro}

Formation of wrinkled zones of thin, hyperelastic stretched sheets has been widely discussed in the literature recently. While the wrinkling of sheets under in-plane compression is a classical buckling problem \cite{Timoshenko:1963,Golubitsky:1984}, the somehow counter-intuitive occurrence of wrinkles under dominantly tensile loading has become intensively discussed around 2000 \cite{Friedl:2000,Cerda:2002,Cerda:2003}. The mechanical background of the phenomena is well described by the variants of the \emph{tension-field theory} \cite{Reissner:1938,Coman:2007}. This approach focuses on the in-plane stress state and neglects the bending stiffness of the sheet. Depending on the $n^-$ number of the negative eigenvalues of the stress tensor, one can distinguish between taut ($n^-=0$), wrinkled ($n^-=1$) and slack ($n^-=2$) zones of the domain. 

Intuitively the wrinkling pattern of a stretched, rectangular sheet clamped at two opposite sides (with the two other sides being free) occurs due to a slight lateral compression caused by the restrained contraction at the clamped ends. Having a sufficiently thin film, even this slight lateral compression is unbearable in a planar state, a buckling occurs and we are left with a wrinkled solution. Tension-field theory is a perfect tool to determine an upper bound for the subdomain affected by wrinkling, but it is unable to predict the wrinkled pattern (number and shape of the wrinkles) itself. The direction of the positive principal stress is the sole information about the shape: it is aligned with the wrinkle crests. In the mathematical point of view the versions of tension-field theory are related to the zero limit in the thickness, thus using them for comparison against experimental data requires extra care. 
To handle the non-vanishing (but still small) thickness of the film and to study the shape of the wrinkled pattern the celebrated Föppl-von Kármán plate theory is used extensively. In the geometric point of view the Föppl-von Kármán model \cite{Karman:1910} is a non-linear theory allowing for finite out-of plane deformations as long as the in plane-strains can be regarded infinitesimal. It predicts the appearance of wrinkles for (presumably perfect) films at a small value of the stretch (much below 1\%). Accordingly, the assumptions underlying the theory are practically valid at this first bifurcation point at which the planar solution loses stability. Following the traditions of bifurcation theory, in our work we will refer the planar solution as \emph{trivial}, however even this solution is affected by lateral contraction (Fig. \ref{Fig:01}). The postcritical branches emanating from the above mentioned bifurcation point contain the wrinkled solutions. These branches are found to be stable and numerical simulations show that the wrinkled pattern preserves stability leaving the trivial solution unstable for any higher value of the stretch.

A recent paper \cite{Healey:2013} suggested to extend the classical Föppl-von Kármán theory into the regime of finite in-plane strains in the case of isotropic (in specific, Saint Venant - Kirchhoff) material. Numerical simulations based on this model suggest that the infinitesimal strain assumption not only results in quantitative errors in the finite strain regime, but the difference between the two models is indeed a \emph{qualitative} one: in the finite strain model the trivial, planar solution regains stability in a second bifurcation point (at some stretch exceeding the critical stretch of the first bifurcation). Furthermore, just a bounded regime of the aspect ratios exhibits wrinkling, \emph{i.e.} for a fixed length sufficiently narrow or wide films would not wrinkle at all. These findings clearly point to an \emph{isola-center bifurcation} in the model. Our paper is devoted to experimentally verify the predictions of the finite-strain model.

Difficulty of such a verification stems from the fact that hardly any material is linearly elastic in the targeted finite strain regime (max. $50\%$ stretch). Papers in the literature tend to present pure numerical simulations \cite{Davidovitch:2011,Puntel:2011,Nayyar:2011,Damil:2014,Taylor:2014}; experimental works either aim to incorporate plastic response \cite{Wong:2006,Nayyar:2014} or compare results measured on elastomers against numerical data based on linear elasticity \cite{Zheng:2009}. In accordance with the general expectation, the stretch needed for well developed wrinkled patterns (approx. $10-25\%$) causes non-negligible adjustment in the material properties; most of the candidate materials for an experimental verification exhibit significant plastification and other, non-linear and non-recoverable phenomena. A well-chosen elastomer (such as polyurethane) is still \emph{hyperelastic} in this range, but it is far-not linear-elastic. We demonstrate, that \emph{prestressing} substantially helps to obtain a nearly linearly elastic material, however, the price of this step must be paid as the material develops non-ignorable \emph{orthotropy}. That is why an orthotropic extension of the original model is essential in case simulations are compared against experimental data. 

Results of the in-plane stress analysis also hint towards emerging orthotropy: in the direction of the stretch the absolute value of the normal stress differs by at least one order of magnitude compared to the cross direction, thus any kind of material adjustment is much more probable in the longitudinal direction. Our experiments affirm that elastomers (in specific: polyurethane) become highly orthotropic during the first extension, after that the material behaves dominantly elastic with negligible variation of the material parameters during more than 10 loading cycles which enables one to carry out repeated elongations of the specimen. We are aware the fact, that restrained contraction (i.e. Poisson-effect) is not the sole reason behind wrinkling, for instance shear deformations (warping) has a contribution, too \cite{Silvestre:2015}. Although this contribution is an interesting experimental challenge, in this paper we focus on emerging orthotropy and the experimental validation of the disappearance of wrinkles. There are a few publication of related experimental results, some of them uses highly elaborated experimental settings (e.g \cite{Davidovitch:2011,Florel:2015}), none of them point out the existence of the second bifurcation point a the resumed flat state of the film. Regarding that lack of observation it is worthy to point out, that even a simple experimental arrangement reproduces the anticipated behavior in a reliable manner due to the well-chosen material and the prestressing. Our findings also point up that the adjustment of the material properties during the first extension requires either extra care from the experimental side or an elaborated model is needed for proper understanding. To the best of our knowledge, this issue is completely overlooked in the literature of thin films.

The paper is organized as follows: to avoid superfluous repetitions we summarize the key findings of \cite{Healey:2013} about the finite strain extension of the Föppl-von Kármán theory in Section \ref{Sect2} and also provide an extension of the model for orthotropic materials. In the following section we investigate our problem applying a viewpoint close to tension-field theory. Section \ref{Sect4} is devoted to the experimental setup and the detailed description of the performed measurements. In Section \ref{Sect5} comparison between numerical and experimental data is presented, finally we draw conclusions in Section \ref{Sect6}.

\section{A model of thin orthotropic plates under finite in-plane strains}
\label{Sect2}

\begin{figure}[!ht]
\centering
\includegraphics[width=0.70\textwidth]{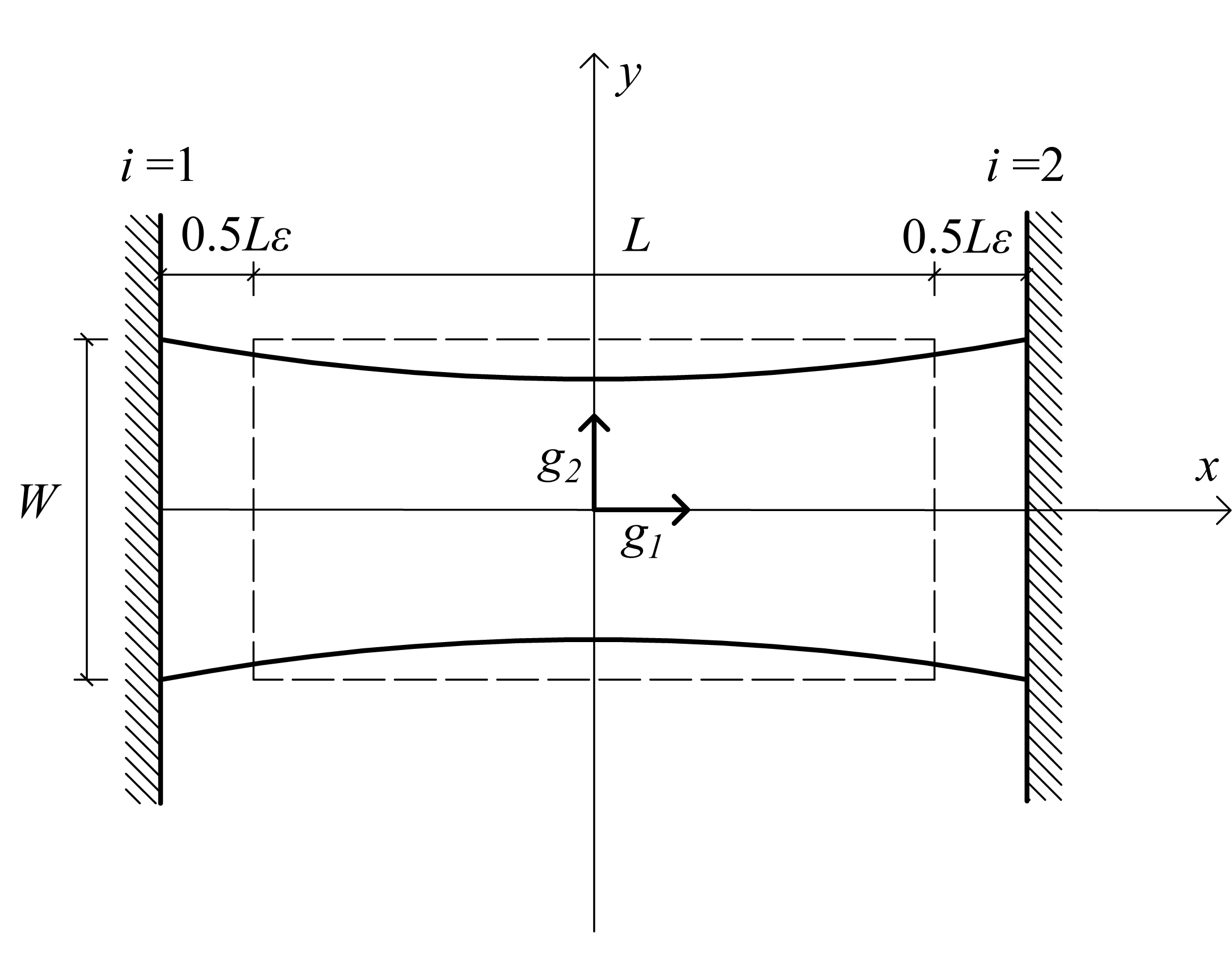}
\caption{The laterally contracted, stretched thin film under $\varepsilon$ stretch. Two opposite sides ($i=1,2$) of the rectangular domain are clamped, the other two are free. The boundary of the domain in the reference configuration with length $L$ and width $W$ is dashed.}
\label{Fig:01}
\end{figure}

We investigate static equilibrium configurations of a thin plate with a constant thickness $h$ occupying a closed domain $\Omega$ in $\xR^3$. The plate is made of a homogeneous material. We use a fixed, orthonormal basis $\{\xg_1,\xg_2,\xg_3\}$, $\xg_1$ and $\xg_2$ span the  plane of the film in the reference configuration (Fig. \ref{Fig:01}). Let us denote the width and the length of the unloaded film by $W$ and $L$, respectively. The \emph{aspect ratio} of the domain is defined as 
\begin{equation}
\label{eq:beta:def}
\beta=\frac{L}{2W}.
\end{equation}

During the deformation the normals of the mid-surface in the reference configuration are mapped to the normals of the deformed mid-surface. Furthermore, only the coupling between the bending and in-plane strains is considered as a (geometrically) non-linear effect. These assumptions enable one to uniquely describe the deformation of the plate by displacements of its middle surface. The $\xu:\Omega\rightarrow \xR^3$ displacement field is given by
\begin{equation}
\xu=
\begin{bmatrix}
u(x,y)\\
v(x,y)\\
w(x,y)\\
\end{bmatrix},
\end{equation}

\noindent where the in-plane displacements $u(x,y)$ and $v(x,y)$ and the out-of-plane deformation $w(x,y)$ are suitably continuously differentiable functions. In the classical Föppl-von Kármán theory the $\Psi(\xu)$ strain energy density of the plate is given as the sum of the energy densities associated with the in-plane deformation ($\Psi_m(\xe)$) and bending ($\Psi_b(\xk)$):
\begin{equation}
\label{eq:eng_den}
\Psi(\xu)=\Psi_m(\xe)+\Psi_b(\xk),
\end{equation}

\noindent where $\xe$ denotes the in-plane strain tensor established by truncating the length-change of an infinitesimal segment in the mid-surface. Furthermore $\xk$ is the linearized bending strain tensor. Using subscripts to denote partial derivatives the linearized strain tensors are formulated as
\begin{eqnarray}
\label{eq:strain_tensors}
\xe(\xu)=\frac{1}{2}
  \begin{bmatrix}
  2u_x+w_x^2      & u_y+v_x+w_xw_y \\
  u_y+v_x+w_xw_y  & 2v_y+w_y^2
  \end{bmatrix}, \\
  \label{eq:lin_bending_strain}
  \hat{\xk}(\xu)=-
  \begin{bmatrix}
  w_{xx}  & w_{xy} \\
  w_{xy}  & w_{yy}
  \end{bmatrix}z=
  -\xk(\xu) z.
\end{eqnarray}

For isotropic materials with a $Y$ modulus of elasticity and $\nu$ Poisson ratio ($0<\nu<0.5$) the theory can be interpreted as a two dimensional approximation of a three dimensional theory \cite{Steigmann:2008} for the Saint-Venant Kirchhoff material: 
\begin{eqnarray}
\label{eq:MembE}
\Psi_m=\frac{Yh}{2(1-\nu^2)}\left[\nu(\Tr{\xe})^2+(1-\nu)\xe\cdot \xe\right],\\
\Psi_b=\frac{Yh^3}{24(1-\nu^2)}\left[\nu(\Tr{\xk})^2+(1-\nu)\xk\cdot \xk\right].
\end{eqnarray}

\noindent We assume a plane-stress problem (i.e. the surfaces of the plate are stress-free) and investigate hard-loading problems, i.e. along subsets $\partial\Omega_i \subset \partial\Omega$, ($i>0$) the displacement is prescribed as
\begin{equation}
\label{eq:bc}
\xu_i=\varepsilon\xu_{0,i}= \varepsilon
\begin{bmatrix}
  u_{0,i} \\
  v_{0,i} \\
  0
  \end{bmatrix},
\end{equation}

\noindent where $u_{0,i}$ and $v_{0,i}$ are given, sufficiently differentiable functions and the scalar $\varepsilon$ is the load parameter. In our case $i=\{1,2\}$ with $u_{0,1}(-L/2,y)=-L/2$, $u_{0,2}(L/2,y)=L/2$ and $v_{0,1}(-L/2,y)=v_{0,2}(L/2,y)=0$ (Fig. \ref{Fig:01}). Assuming no other external loads the $I(\xu)$ potential energy of the system is simply
\begin{equation}
\label{eq:pot_e}
I(\xu)=\int_{\Omega}\Psi \textnormal{d}\Omega.
\end{equation}

We seek solutions that minimize the functional in (\ref{eq:pot_e}) subject to the boundary conditions (\ref{eq:bc}). Extension of the model to the finite strain regime is carried out by using the 
\begin{multline}
\label{eq:lag_strain}
\xE(\xu)=\\
\frac{1}{2}
\begin{bmatrix}
2u_x+u_x^2+v_x^2+w_x^2       & u_y+v_x+u_xu_y+v_xv_y+w_xw_y \\
u_y+v_x+u_xu_y+v_xv_y+w_xw_y & 2v_y+u_y^2+v_y^2+w_y^2
\end{bmatrix}
\end{multline}

\noindent Green-Lagrangian strain tensor instead of the truncated strain tensor $\xe$ in eq. (\ref{eq:MembE}). The energy density is obtained as 
\begin{equation}
\label{eq:pot_e2}
\Psi(\xu)=\Psi_m(\xE)+\Psi_b(\xk).
\end{equation}

\noindent For details and mathematical justification of this (not straightforward) substitution we refer to the above citepd paper\cite{Healey:2013}. The first variations of (\ref{eq:pot_e}) with respect to the components of the displacement-field $\xu$ deliver the equilibrium (Euler-Lagrange) equations:
\begin{eqnarray}
\label{eq:EL-1}
\nabla\cdot\left[\left(\xI+\xg_1\otimes\nabla u+\xg_2\otimes\nabla v\right)\xN\right]=\x0, \\
\label{eq:EL-2}
h^2\Delta^2w-\nabla\cdot\left(\xN\nabla w\right)=0,
\end{eqnarray}

\noindent where $\xI$ is the ($2 \times 2$) unit tensor and $\Delta(.)$ and $\Delta^2(.)$ denote the Laplace and biharmonic operators, respectively. The stress tensor denoted by $\xN$ in the equilibrium equations is the second Piola-Kirchhoff stress associated with the membrane behavior, in specific:
\begin{equation}
\label{eq:P-K:iso}
\xN=\frac{\textnormal{d}\psi_m}{\textnormal{d}\xE}=\frac{Yh}{(1-\nu^2)}\left[\nu\Tr(\xE)\xI+(1-\nu)\xE\right].
\end{equation}

The potential energy functional for films made of orthotropic material is obtained in a similar manner to the isotropic case \cite{Libai:1998}. We assume that the principal directions of the material orthotropy are being aligned with the $x$ and $y$ coordinate axes. Let $Y_0$ and $Y_{90}$ denote the moduli of elasticity in directions $x$ and $y$, respectively. We define the ratio 
\begin{equation}
\label{eq:def_r}
r=\frac{Y_{90}}{Y_0}
\end{equation}

\noindent to denote the \emph{degree of orthotropy}. Furthermore, symmetry conditions on the elastic stiffness tensor imply that the Poisson ratios in the two principal directions fulfill $\nu_{[xy]}=r\nu_{[yx]}$ \cite{Howell:2009}. We express the shear modulus $\mu$ as
\begin{equation}
\label{eq:def_q}
\mu=qY_0,
\end{equation}

\noindent with $q$ being a positive scalar. Let us write the fourth order material elastic stiffness tensor $\xC$ of a two dimensional, orthotropic medium in the following way:
\begin{equation}
\xC=\frac{Y_{0}}{1-r\nu_{[xy]}^2}\widetilde{\xC},
\end{equation}

\noindent where the nonzero elements of the fourth order tensor $\widetilde{\xC}$ solely depend on the non-dimensional material parameters: $\widetilde{C}_{1111}=1$, $\widetilde{C}_{1122}=\widetilde{C}_{2211}=r\nu_{[xy]}$, $\widetilde{C}_{2222}=r$ and finally $\widetilde{C}_{1212}=q(r\nu_{[xy]}^2-1)$. To obtain the energy densities $\Psi_m$ and $\Psi_b$ for orthotropic films we use the linearized bending tensor (\ref{eq:lin_bending_strain}), the finite strain in-plane tensor from (\ref{eq:lag_strain}) and integrate through the thickness of the film:
\begin{eqnarray}
\label{eq:ed_ortho_membarne}
\Psi_m(\xE)=\frac{1}{2}\int^{h/2}_{-h/2}\left\{\xE\cdot\xC\cdot\xE\right\}\textnormal{d}z=\frac{1}{2}\frac{Y_{0}h}{1-r\nu_{[xy]}^2}\xE\cdot\widetilde{\xC}\cdot\xE,\\
\label{eq:ed_ortho_bending}
\Psi_b(\xk)=\frac{1}{2}\int^{h/2}_{-h/2}\left\{\xk\cdot\xC\cdot\xk\right\}\textnormal{d}z=\frac{1}{24}\frac{Y_{0}h^3}{1-r\nu_{[xy]}^2}\xk\cdot\widetilde{\xC}\cdot\xk.
\end{eqnarray}

The potential energy of the enitre domain follows by substituting (\ref{eq:ed_ortho_membarne}) and (\ref{eq:ed_ortho_bending}) into eq.(\ref{eq:pot_e2}). After rescaling (i.e. multiplying by $24(1-r\nu_{[xy]}^2)/(Y_0h^3)$) and introducing $\kappa=h^{-2}$ we obtain the energy functional
\begin{equation}
\begin{split}
\label{eq:orto_e}
I(\xu)&=\int_{\Omega}\left\{\Psi_m(\xE)+\Psi_b(\xk)\right\}\textnormal{d}\Omega\\
&=\int_{\Omega}\left\{ 12\kappa\xE\cdot\widetilde{\xC}\cdot\xE+\xk\cdot\widetilde{\xC}\cdot\xk \right\}\textnormal{d}\Omega.
\end{split}
\end{equation}

The second Piola-Kirchhoff stress tensor arising from membrane strains can be formulated as 
\begin{equation}
\label{eq:P-K:ortho}
\xN=\frac{\textnormal{d}\psi_m}{\textnormal{d}\xE}=
12\begin{bmatrix}
  \xE_{11}+r\nu_{[xy]}\xE_{22}  & 2q(1-r\nu_{[xy]}^2)\xE_{12} \\
   2q(1-r\nu_{[xy]}^2)\xE_{12}  & r\xE_{22}+r\nu_{[xy]}\xE_{11}
\end{bmatrix}.
\end{equation}

\noindent Nevertheless, for $r=1$ and $q=0.5(1+\nu)^{-1}$ this is identical to the rescaled version of the isotropic case from eq.(\ref{eq:P-K:iso}). The Euler-Lagrange equations associated with the energy functional in (\ref{eq:orto_e}) are formally identical to equations (\ref{eq:EL-1}-\ref{eq:EL-2}), nevertheless the stress $\xN$ is given by (\ref{eq:P-K:ortho}) in this case.

The applied numerical scheme (in fact, a Ritz method) aims to approximate the minima of the functional (\ref{eq:orto_e}) subject to the boundary conditions in eq.(\ref{eq:bc}) by finite element discretization. Observe, that the material parameters involved in the problem are the dimensionless $r$, $q$ and $\nu_{[xy]}$. We use a regular rectangular mesh to establish the finite elements \cite{Reddy:2008} with the usual $C_0$ basis functions to approximate $u$ and $v$ and $C_1$ basis functions for $w$. Each internal node of the discretization represents six unknowns. The compiled system of nonlinear algebraic equations are the discretized Euler-Lagrange equations associated with eq.(\ref{eq:orto_e}). 

To investigate stable, wrinkled solutions along non-trivial equilibrium branches classical arch-length continuation is a perfect numerical tool \cite{Healey:2013}. Since in our work we target to show that planar solutions along the trivial equilibrium branch regain stability the investigation of the trivial solution is sufficient. We aim to demonstrate that for sufficiently large, fixed $\beta$ there exist $0<\varepsilon_{cr,1}<\varepsilon_{cr,2}<\infty$ such that the trivial branch is unstable iff $\varepsilon_{cr,1}<\varepsilon<\varepsilon_{cr,2}$ and it is stable otherwise. To reach our goal we evaluate the minimal eigenvalue $\lambda_{\min}$ of the Jacobian (\emph{i.e.} the second variation of the discrete approximation of functional (\ref{eq:orto_e}), as it is usual for Ritz methods) at solutions along the trivial equilibrium path. The trivial, planar ($w\equiv 0$) solution is unstable for $\lambda_{\min}<0$, otherwise it is considered to be stable. In case we needed to pick up a solution from the non-trivial branch we fixed all the parameters in the problem, applied a random perturbation for the trivial, planar solution and used Broyden's method to detect a wrinkled solution.

\section{Investigation of the parameter space: a stress analysis}
\label{Sect3}

For given material parameters (\emph{i.e.} $r$, $q$ and $\nu$ is fixed) our problem consists of three bifurcation parameters: the $h$ thickness of the film, the $\beta$ aspect ratio of the domain and the applied stretch $\varepsilon$. The already citepd paper \cite{Healey:2013} presents a lemma, which leads to the conclusion, that a necessary condition for a bifurcation from the trivial, planar state (\emph{i.e.} $w\neq 0$ for some $(x,y)\in\Omega$) is the violation of the non-negativity of the second Piola-Kirchhoff stress tensor computed for the trivial solution ($w\equiv 0$). This statement remains valid for an orthotropic material. For a planar solution the left-hand side of (\ref{eq:EL-2}) is identically zero for any $h$ and $\xN$, thus the solution of (\ref{eq:EL-1}) is independent of the film thickness. We arrive at the conclusion that the negativity of $\xN$ (computed for a trivial solution) can be used to determine a point-set of possible bifurcation points in the $\beta$-$\varepsilon$ quarter for any positive values of $h$.  

\begin{landscape} 
\begin{figure}
\centering
\includegraphics[width=1.35\textwidth]{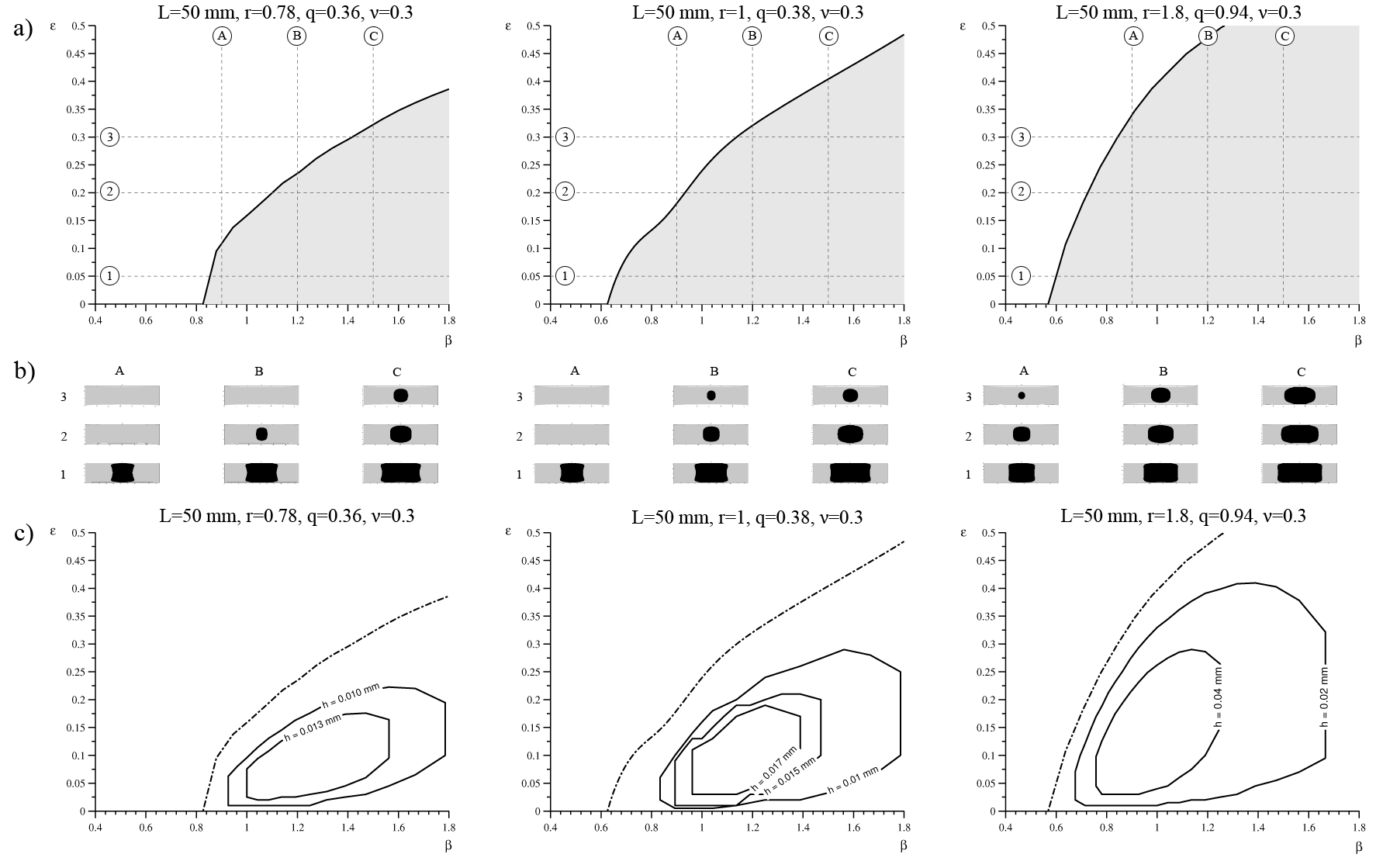}
\caption{Numerical investigation of the parameter-space. a) The shaded point-set below the black curve refers to $\beta-\varepsilon$ pairs at which the stress tensor $\xN$ computed for the trivial solution is negative (for any thickness $h$) at three different degrees of orthotropy (see text for details) b) Domain of the film shaded according to the negativity of the stress tensor: black (gray) refers to $n^{-}=1$ ($n^{-}=0$, respectively) negative eigenvalue of $\xN$ at the designated points of the graphs above c) bifurcations: for a fixed $h$ the smallest eigenvalue of the Jacobian ($\lambda_{\min}<0$) determines a point set surrounded by a closed curve, inside the curve the trivial solution is unstable, the stable solution is wrinkled. The dashed-dotted line is the edge of the zone with negative stress (as in graphs in the a) part of the figure above).}
\label{Fig:02}
\end{figure} 
\end{landscape}

Negativity of the stress tensor marks the admissible locations of the two bifurcation points $\varepsilon_{cr,1}$ and $\varepsilon_{cr,2}$ in the parameter space (Fig. \ref{Fig:02}). This approach reveals several, so far unrevealed facts: in the $\varepsilon$ direction as $\varepsilon\rightarrow\infty$ the stress tensor becomes positive, which means that even in case $h\rightarrow 0$ the $\varepsilon_{cr,2}$ value for the second bifurcation point (this is the one, where the trivial solution regains stability) remains finite. This observation is not purely numerical: let us consider our governing equations for a given domain (\emph{i.e.} $\beta$ is fixed) with an arbitrary large stretch. To make our boundary value problem homogeneous we apply the following change of variables:
\begin{equation}
\label{eq:subs}
u(x,y)=\varepsilon x + \bar{u}(x,y).
\end{equation}

\noindent Substituting (\ref{eq:subs}) into the Green-Lagrangian strain tensor in eq. (\ref{eq:lag_strain}) and using eq. (\ref{eq:P-K:ortho}) the two components of the (\ref{eq:EL-1}) equilibrium equation can be arranged as follows:
\begin{equation}
\begin{split}
\label{eq:equ:limit}
\left(18\bar{u}_{xx}+6r\nu\bar{u}_{yy}-12qr\nu^2\bar{u}_{yy}+12q\bar{u}_{yy}\right)\varepsilon^2 + \\
+f_1\left(.\right)\varepsilon+g_1\left(.\right)=0,
\end{split}
\end{equation}

\begin{equation}
\label{eq:equ:limit2}
\left(6v_{xx}+6r\nu v_{yy}\right)\varepsilon^2+f_2\left(.\right)\varepsilon+g_2\left(.\right)=0,
\end{equation}

\noindent where $f_1(.)$, $f_2(.)$, $g_1(.)$ and $g_2(.)$ all depend on the first and second partial derivatives of $\bar{u}$ and $v$. Division of both equations by $\varepsilon^2$ and letting $\varepsilon\rightarrow\infty$ leads to two, uncoupled equations:
\begin{eqnarray}
\label{eq:equ:limit3}
\left(18\bar{u}_{xx}+6r\nu\bar{u}_{yy}-12qr\nu^2\bar{u}_{yy}+12q\bar{u}_{yy}\right)=0,\\
\label{eq:equ:limit4}
\left(6v_{xx}+6r\nu v_{yy}\right)=0.
\end{eqnarray}

Taking into account the boundary conditions of the clamped sheet and the reflectional symmetry along the coordinate axes ($\bar{u}=0$ along $x=0$ and $v=0$ along $y=0$) solutions of the PDEs in (\ref{eq:equ:limit3}) and (\ref{eq:equ:limit4}) can be written in a closed form:
\begin{eqnarray}
\label{eq:equ:sol1}
\bar{u}(x,y)=2b\cosh\left(ay\right)\sin\left(a\frac{\sqrt{6q-6qr\nu^2+3r\nu}x}{3}\right),\\
\label{eq:equ:sol2}
v(x,y)=-2c\sinh\left(dy\right)\cos\left(d\sqrt{r\nu}x\right),
\end{eqnarray}

\noindent where $a$, $b$, $c$ and $d$ are appropriate real numbers. In specific, $a$ and $d$ should be chosen to fulfill the boundary conditions along the clamped sides. Since none of these constants are $O(\varepsilon)$ from eqs. (\ref{eq:lag_strain}) and (\ref{eq:P-K:ortho}) we obtain:
\begin{equation}
\label{eq:P-K:ortho_limit}
\lim_{\varepsilon\rightarrow\infty}\frac{1}{\varepsilon^2}\xN=
6\begin{bmatrix}
   1  & 0 \\
   0 & r\nu
\end{bmatrix}>0.
\end{equation}
Computation of the negative eigenvalue of the stress tensor revealed, that at a fixed finite stretch the negative eigenvalue persists for arbitrary high values of $\beta$ (Fig. \ref{Fig:02} a)); however, the maximal compressive stress gradually decreases as $\beta$ tends to infinity. In other words elongated, flat sheets are sensitive in material and geometrical imperfections as compression is presented to advocate wrinkling, hence bending rigidity need to be sufficient to obstacle it. In Figure 2 we present computational results for three different degrees of orthotropy: the middle column belongs to the isotropic material, the left column is plotted for a small orthotropy typical for the unloaded polyurethane sheet and the right column shows the results for a highly orthotropic material (\emph{i.e.} the polyurethane sheet after prestressing). As we mentioned, negativity of $\xN$ is necessary, but not sufficient condition to have a bifurcation from the trivial solution. For fixed values of $h$ we also computed the closed curves of bifurcation points (Fig. \ref{Fig:02} c)), nevertheless, these non-intersecting curves are shrinking as $h$ is increased and over a critical thickness there is no wrinkling at all.

\section{Experimental}
\label{Sect4}

\begin{figure}[!ht]
\centering
\subfloat[$\varepsilon$ = 0.00]{
   \includegraphics[width=0.65\textwidth]{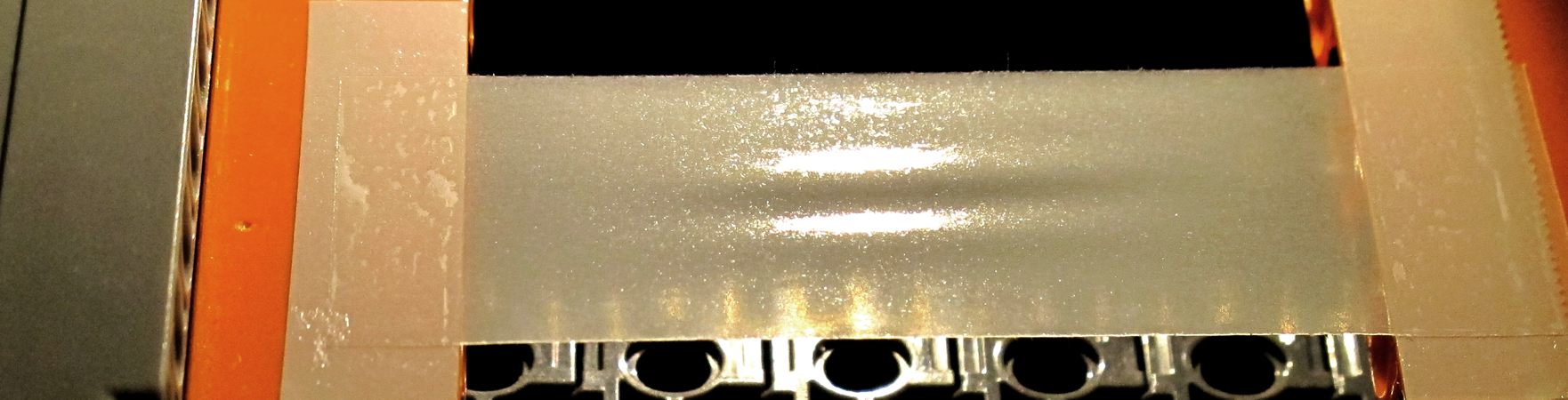}
}\hfill
\subfloat[$\varepsilon$ = 0.04]{
   \includegraphics[width=0.65\textwidth]{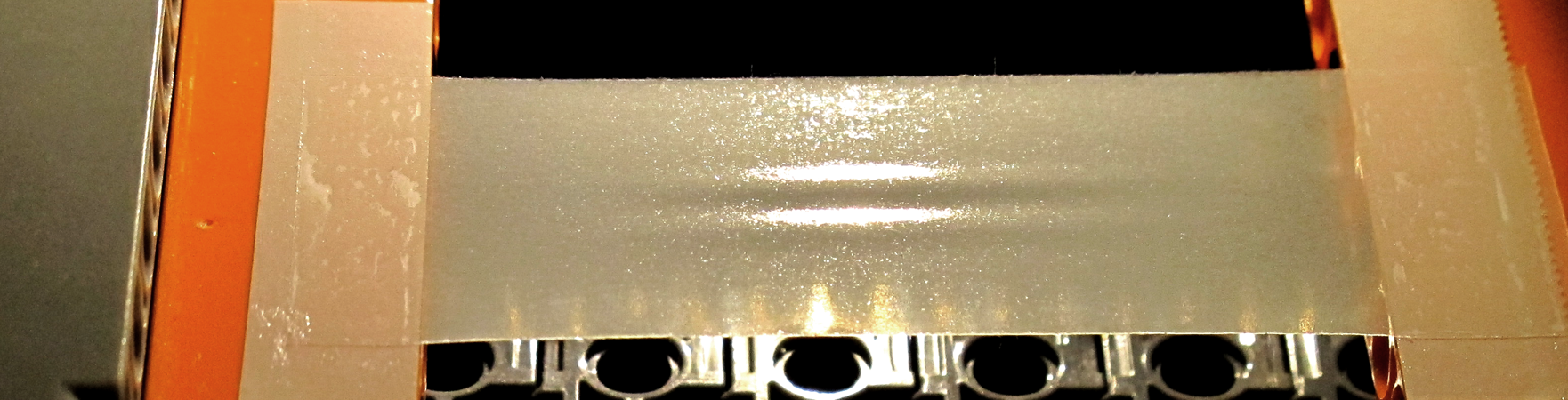}
} \hfill
\subfloat[$\varepsilon$ = 0.13]{%
   \includegraphics[width=0.65\textwidth]{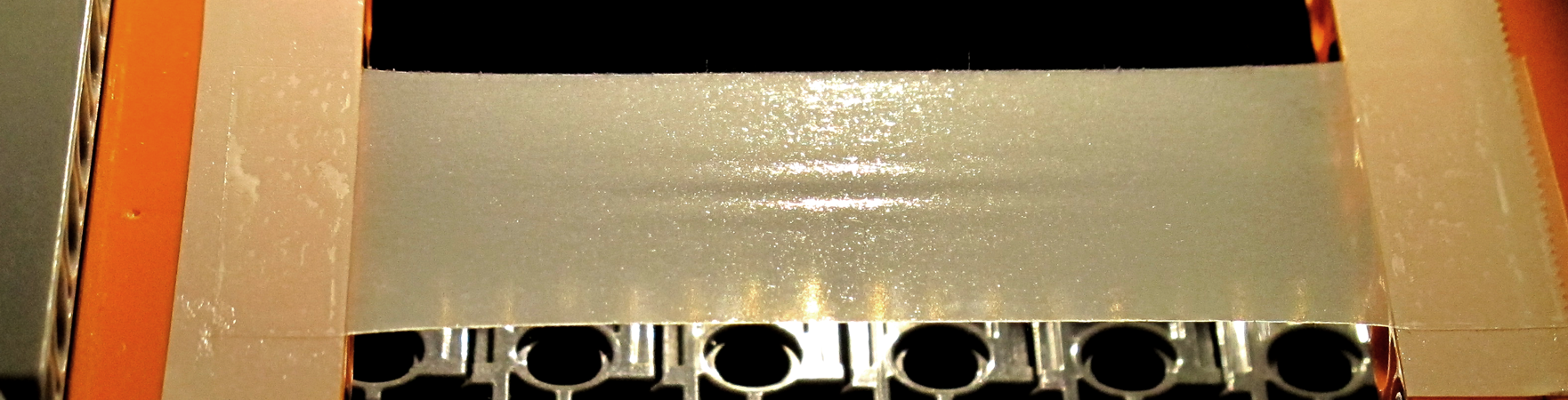}
} \hfill
\subfloat[$\varepsilon$ = 0.22]{%
   \includegraphics[width=0.65\textwidth]{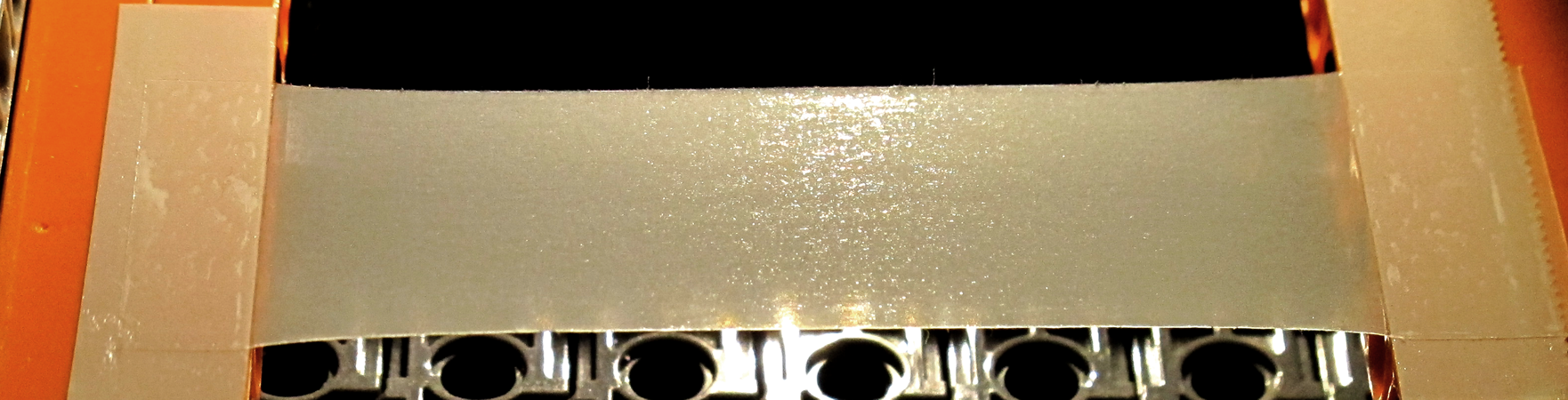}
}
\caption{Increasing of the stretch leads to the disappearance of wrinkles. L = 50 mm, W = 25 mm, $\beta = 1$.}
\label{Fig:03}
\end{figure}

The experimental investigation of wrinkling is not straightforward:  the realization of the idealized boundary conditions and application of the proper loading might call for a special, unique equipment  \cite{Jenkins:1998,Geminard:2004}. 
As our case is very close to a traditional pulling test and we model a displacement controlled process, a direct measurement of the stresses along the clamped boundary of the film is not necessary. Hence we determined the material properties separately and examined wrinkling with a simple machine equipped with a servo engine. The speed of the extension was fixed at $120$ mm/min in all tests done with this machine.  The occurrence and disappearance of the wrinkles were observed visually, in addition we photographed the specimen in small-angle lighting \cite{Geminard:2004} (Fig. \ref{Fig:03}) and recorded videos (provided as supplementary data). Visual observation does not provide an exact critical stretch at disappearance ($\varepsilon_{cr,2}$). We determined $\varepsilon_{21}$, at which the wrinkles were definitely visible and a $\varepsilon_{22}$ with a visibly plane specimen without any shades. Of course, we expect $\varepsilon_{21}<\varepsilon_{cr,2}<\varepsilon_{22}$. More accurate observation technologies (\emph{e.g.} optical measuring or scanning the middle section of the specimen) would close the gap between the measured lower and upper bounds. We took $\varepsilon_2=0.5(\varepsilon_{21}+\varepsilon_{22})$.

There is a huge range of elastomer films available. We tested several products used as health-care tapes and carried out our experiments on Hydrofilm Roll manufactured by Paul Hartmann AG. The used sheets have a $20$ \SI{}{\micro\metre} nominal thickness and they are covered by an adhesive layer on one side. As it is typical among polymers \cite{Ward:1997}, the unloaded material is slightly orthotropic due to the manufacturing process. In addition, as we already pointed out, the first extension (let us call it \emph{prestressing}) results in a significant change in the material properties.

\begin{table}[!ht]
	\centering
	\subfloat{%
		\begin{tabular}{l c}
			\hline
			\multicolumn{2}{c}{Before prestressing} \\
			\hline
			Y$_0$ & 48.45 N/mm$^2$\\
			Y$_{90}$ & 37.75 N/mm$^2$ \\
			Y$_{45}$ & 44.23 N/mm$^2$ \\
			G & 17.41 N/mm$^2$ \\
			r & \textbf{0.78} \\
			q & \textbf{0.36} \\
			$\nu$ & 0.3 \\
			\hline
		\end{tabular}
	}\hspace{1cm}
	\subfloat{%
		\begin{tabular}{l c}
			\hline
			\multicolumn{2}{c}{After prestressing} \\
			\hline
			Y$_0$ & 15.36 N/mm$^2$\\
			Y$_{90}$ & 27.69 N/mm$^2$ \\
			Y$_{45}$ & 28.66 N/mm$^2$ \\
			G & 14.55 N/mm$^2$ \\
			r & \textbf{1.80} \\
			q & \textbf{0.94} \\
			$\nu$ & 0.3 \\
			\hline
		\end{tabular}
	}
	\caption{Material parameters measured before and after prestressing. Observe the remarkable shift in the degree of orthotropy. The measured values used to compute the material parameters are provided as supplementary data (MaterialParameters.pdf).}
	\label{Table:01}
\end{table}

The following loading cycles turned out to have negligible further effect on the material features, up to 8-10 loading cycles the material can be regarded as hyperelastic, what is more, it is rather close to be linear elastic. We fixed the applied $\varepsilon=0.66$ prestretch in all measurements and experiments, since we expected the wrinkles to disappear below this threshold.

The material properties were determined for both the unloaded and the prestressed material in three test series. In the first series we used a Zwick Z150 material testing machine equipped with special grips (type: Zwick 9103 10) developed for technical membranes. The testing machine detected the force-displacement diagram during the displacement controlled pull test in $0.01$ mm steps. In all series we measured 5 specimens with a width of $W=30$ mm and length of $L=25$ mm cut out parallel, in 45 degrees and perpendicular to the long direction. Results of these three directions provided all the needed material parameters ($Y_0$, $Y_{45}$, $Y_{90}$, $\nu_{[xy]}$, $r$, $q$) \cite{Lempriere:1968,Cho:2001}. In the second and third test series (let us call them control tests) we used a Zwick Z020 material testing machine and a videoextensometer (Messphysik ME 46 full image videoextensometer), and determined the force-displacement diagram in $0.002$ mm intervals. The three testing series resulted in very close outcomes and the variation of the material parameters were also low. The best fit for the measured data for the unloaded material resulted in $r=0.78$ and $q=0.36$, while for the prestressed case the strong direction of orthotropy interchanged as we measured $r=1.80$ and $q=0.94$ with negligible change in the Poisson ratio $\nu=\nu_{[xy]}$ (Table \ref{Table:01}).

\section{Results and discussion}
\label{Sect5}

\begin{figure}[!ht]
\centering
\includegraphics[width=0.65\textwidth]{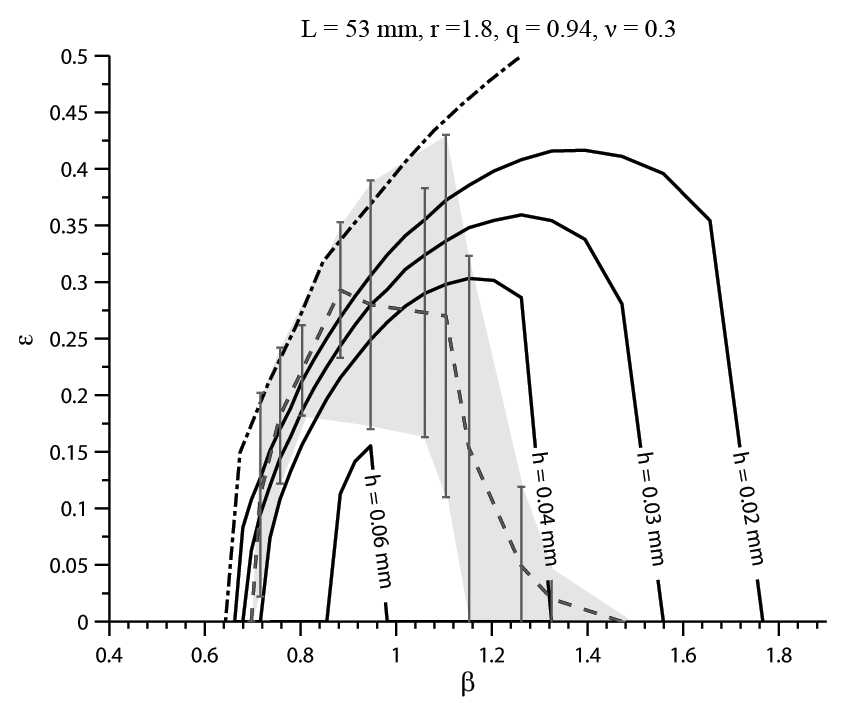}
\caption{Comparison of numerical simulations against experimental data. The dot dashed line set out the border for negative PK stress (as it is given in Fig. \ref{Fig:02}). Numerical values for the $\varepsilon_{cr,2}$ critical points at which wrinkles disappear are plotted with solid lines for several thickness values. The dashed line stands for the experimental result, observe the good agreement between the mean of the experiments and the computed results for $h=40$ \SI{}{\micro\metre}. The measured values are provided as supplementary data (Measured.pdf).}
\label{Fig:05}
\end{figure} 

As we pointed out, disappearance of wrinkling of stretched films were predicted based on numerical simulations \cite{Healey:2013}. Some other numerical studies \cite{Nayyar:2014} also point to such a phenomenon. On the other hand, experimental works lack to confirm this prediction. Our experiments clearly demonstrated, that in accordance with the prediction, wrinkles disappear over a critical stretch. Thermoplastic polymers (such as polyethylene and polypropylene) used in the experimental works are inferior choices for such a verification due to their dominantly plastic behavior. In spite of this difficulty, in our earlier work \cite{Feher:2014} we reported about disappearance of wrinkling in experiments with polypropylene sheets. That was a \emph{qualitative} agreement, a quantitative comparison would be meaningful against simulations of plastic constitutive relations. Such a relation has been published recently \cite{Nayyar:2014}, however that work completely lacks the investigation of emerging orthotropy during the loading process. This change in the orthotropy parameters is so significant that for some aspect ratios no wrinkling appears at all during the first extension of the specimen, but in the successive cycles it forms repeatedly (watch the video file provided as supplementary data). This observation also supports our methodology, namely, that we carry out our investigation on prestressed elastomers. After prestressing the material is close to be linear elastic, thus in this case \emph{quantitative comparison} against numerical simulations is possible.

For numerical simulations we used the average of the measured material parameters mentioned at the end of the previous section. Before prestress the free distance between the clamped ends was adjusted to $50$ mm. During the prestress approximately $3$ mm plastic (unrecoverable) extension of the film was observable (regardless of the width of the film), thus to keep consistence with the experiments in the numerical simulations we took $L=53$ mm into account. Since the thickness of the film contains some uncertainty due to the presence of the adhesive layer we plotted the numerical results for a reasonable range (Fig. \ref{Fig:05}). While numerical results for $h=20$ \SI{}{\micro\metre} exceeds the curve of the experimental results, then the graph for $h=60$ \SI{}{\micro\metre} is completely inside the curve of the measurements, the agreement with $h=40$ \SI{}{\micro\metre} is convincing. For such a high degree of orthotropy the first bifurcation point (appearance of wrinkles, $\varepsilon_{cr,1}$) is located very close to $\varepsilon=0$ (see Fig. \ref{Fig:02}). These points are omitted from Fig. \ref{Fig:05}, but the very rapid appearance of wrinkles in the experiments is in accordance with the simulations. As we mentioned earlier, higher values of $\beta$ are considered to be more imperfection sensitive, which is also confirmed by our experiments: observe the increasing variation between $0.8<\beta<1.2$. It also explains the slightly worse accordance between numerical and experimental data around $\beta=1.2$.

\begin{figure}[!ht]
\centering
\includegraphics[width=0.68\textwidth]{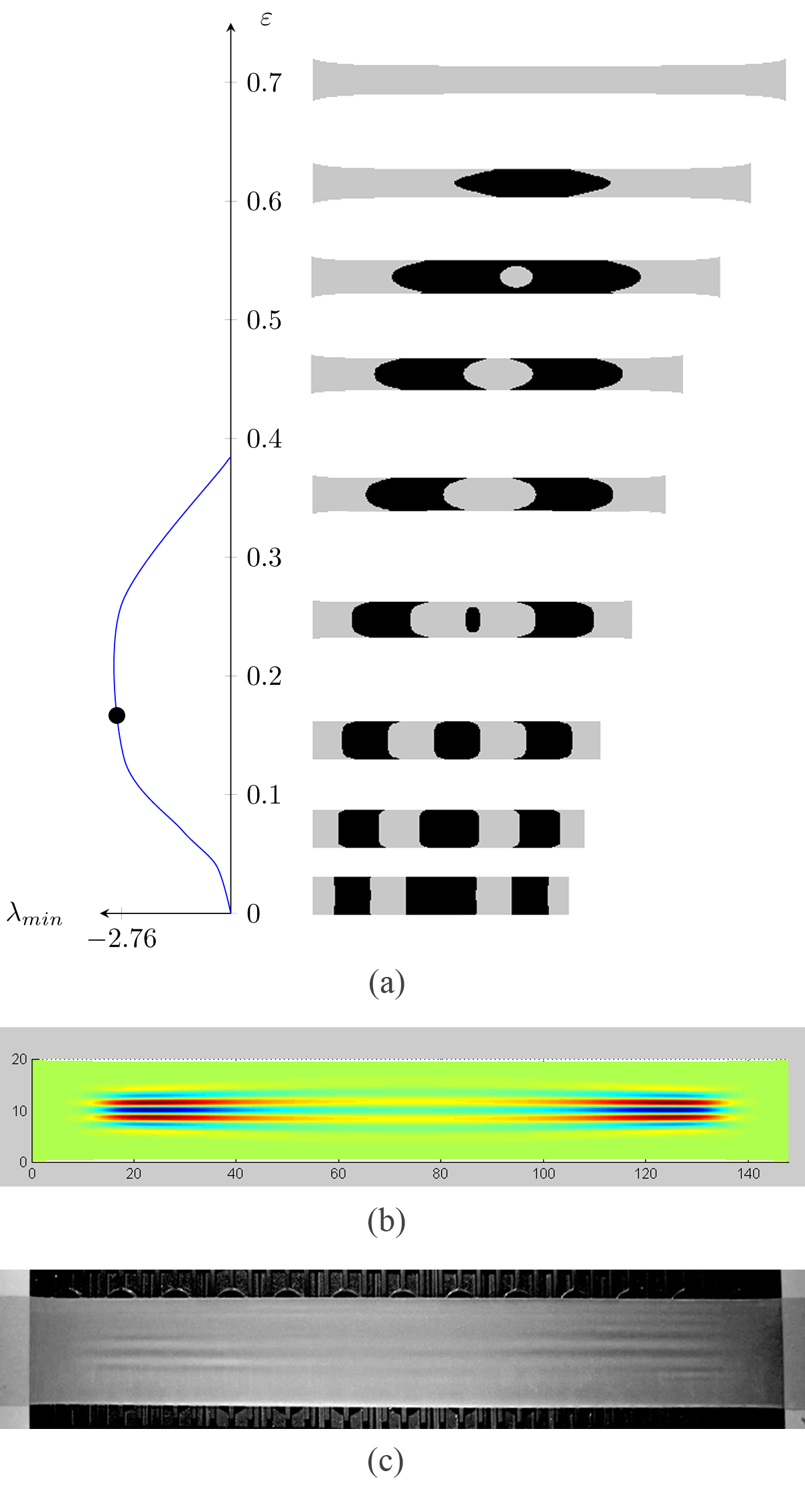}
\caption{For high aspect ratios at the bifurcation points the compressed zones are disconnected. a) At a fixed domain ($W = 20\;$mm, $L = 130\; $mm, $\beta=3.25$, $h=15$ \SI{}{\micro\metre} polyurethane film) the two bifurcation points are located at $\varepsilon_{cr,1} \approx 0.05$ and $\varepsilon_{cr,2}\approx 0.39$ Observe the fusion of compressed zones as the stretch is increased, b) A computed wrinkled solution at $\varepsilon=0.15$ (denoted by black dot in the $\lambda_{\min}-\varepsilon$ graph), c) experimental evidence for the computed arrangement of wrinkles at $\varepsilon=0.15$.}
\label{Fig:06}
\end{figure} 

As we pointed out in Section \ref{Sect3}, compression vanishes for high values of the stretch, but not for high values of the aspect ratios. Elongated sheets (in the trivial, planar state) with $\beta>1.56$ exhibited two, disjoint sets of compressed regions (Fig. \ref{Fig:06}) in the simulations. Observe that for instance at $\beta=3.25$ and $h=0.15$ \SI{}{\micro\metre} the compressed regions are disjoint for the stretch values at which the planar solution is unstable ($0.05<\varepsilon<0.39$, Fig. \ref{Fig:06} a)). Due to the disjoint sets of compressed points in the trivial solution we expected complicated wrinkled patterns for very thin and long, elongated sheets. We present such a solution in the b) and c) parts of Fig. \ref{Fig:06}. Observe, that there are two, well separated, highly wrinkled zones formed both in the simulation and the experiment. These finding are in good agreement with numerically predicted arrangement of wrinkles for such a high aspect ratio \cite{Huang:2015}.

\section{Conclusions}
\label{Sect6}

In our paper we reported about successful experiments to validate a prediction of the Föppl-von Kármán theory extended to finite in-plane strains namely the disappearance of the wrinkled pattern of clamped, thin sheets under uniform tension. Our work not only provide experimental proof about the second bifurcation point along the trivial equilibrium path, at which the planar solution regains stability, but it also confirms that the aspect ratio of the rectangular domain significantly affects the location of the bifurcation points. Our findings support the numerically predicted isola-center bifurcation in the finite strain model. 

To cope with the difficulty of the substantial stretch needed for wrinkled patterns we applied prestress on a well-chosen elastomer (polyurethane) to gain an essentially linearly elastic material behavior. A significant orthotropy emerges during prestressing, according to this observation we extended the originally isotropic model to orthotropic films. Numerical computations based on the orthotropic constitutive behaviour agree well with the experimental results as long as plastic effects are negligible. 

In this paper we focused on incorporation of a realistic, but in the same time uncomplicated constitutive law into a geometrically nonlinear model. It suggests a problem for further research, namely the application of wrinkled patterns to draw conclusions about the constitutive law of the material (in specific about the degree of orthotropy) as well as incorporating the adjustment of the material during the prestress via some well-chosen damage propagation approach, for instance the Mullins-effect.

\section*{Acknowledgment}
We thank \emph{K\'aroly P. Juh\'asz} and \emph{Ott\'o Sebesty\'en} for their valuable help in the experimental work. The Zwick Z150 material testing machine were provided by the T\'AMOP 4.2.1/B-09/1/KMR-2010-0002 grant. We are also grateful for \emph{Marianna Hal\'asz} and  \emph{G\'abor Szeb\'enyi}  for helping us in the control measurements with the videoextensometer. This paper was supported by the János Bolyai Research Scholarship of the Hungarian Academy of Sciences [SA].

\bibliographystyle{plain}
\bibliography{References}
\end{document}